\documentstyle[aps,prl,multicol,amssymb,epsf]{revtex}

\begin{document}
\author{Daniel Jonathan and Martin B. Plenio}
\title{Entanglement-assisted local manipulation of pure quantum states}
\address{Blackett Laboratory, Imperial College, London SW7 2BZ, United Kingdom.}
\date{\today}
\draft
\maketitle

\begin{abstract}
We demonstrate that local transformations on a composite quantum system
can be enhanced in the presence of certain entangled states. These extra
states act much like {\em catalysts} in a chemical reaction: they allow
otherwise impossible local transformations to be realised, without being
consumed in any way. In particular, we show that this effect can considerably 
improve the efficiency of entanglement concentration procedures for finite 
states.
\end{abstract}

\begin{multicols}{2}
The rapid development of quantum information processing in recent years has
led us to view quantum-mechanical entanglement as a useful physical
resource \cite{Plenio1}. As with any such resource, there arises naturally
the question of how it can be quantified and manipulated. Attempts have
been made to find meaningful measures of entanglement
\cite{Bennett96,Bennett96a,Bennett96b,Vedral1,Vedral2}, and also to uncover 
the fundamental laws of its behaviour under local quantum operations 
and classical communication (LQCC)
\cite{Bennett96,Bennett96a,Bennett96b,Vedral1,Vedral2,Lo1,Vidal2,Nielsen,Vidal1,Jonathan1,Hardy1}.
These laws are fundamentally and also practically important, since many
applications of quantum information processing involve spatially separated
parties who must manipulate an entangled state without performing joint
operations. In this context, it is generally assumed that entanglement may
be used to perform useful tasks only if it is consumed in whole or in 
part. Indeed, this is implicit in the common-sense notion of a ``resource''.

In this Letter we demonstrate that entanglement is, in fact, a stranger kind
of resource, one that can be used without being consumed at all. More
precisely, we show that the mere presence of an entangled state can allow
distant parties to realise local transformations that would otherwise be
impossible, or less efficient. Our idea is best introduced by the following
situation, illustrated in Fig. 1. Imagine that Alice and Bob share a
finite-dimensional entangled state $|\psi_1\rangle$ of two particles, which
they would like to convert, using only LQCC, into the state
$|\psi_2\rangle$. For some choices of $|\psi_1\rangle$ and $|\psi_2\rangle$
there exists a local protocol that accomplishes this task with certainty
\cite{Nielsen}, but for others it can only be done probabilistically, with
some maximum probability $p_{max}<1$ \cite{Vidal1}. Assume the latter is
the case, as indicated by the crossed arrow in the upper part of Fig. 1.
Now suppose that an ``entanglement banker'', let us call him Scrooge, agrees
to lend Alice and Bob another entangled pair of particles $|\phi\rangle$,
under the condition that {\em exactly} the same state must be returned to
him later on. Given this additional state, will Alice and Bob be able to
transform $|\psi_1\rangle$ into $|\psi_2\rangle$ and still return the state
$|\phi\rangle$ to Scrooge? We suggest to call a transformation of this
kind, which uses intermediate entanglement without consuming it, an {\em
entanglement-assisted local transformation}, abbreviated by ELQCC.
The possible existence of such a class of transformations 
has been conjectured by Popescu \cite{Sandu1} (see also \cite{Sandu2}).

\begin{figure}[hbt]
\begin{center}
\leavevmode
\epsfxsize=8cm
\epsfbox{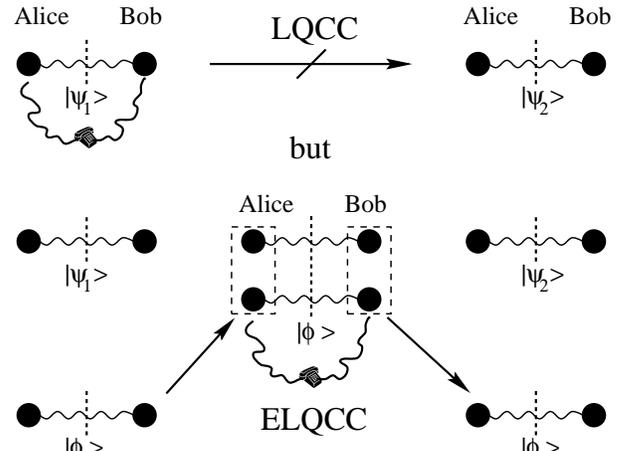}
\caption{\narrowtext Alice and Bob share a finite bipartite system in state
$|\psi_1\rangle$. Using only LQCC they are not able to convert this state
into $|\psi_2\rangle$ with certainty. However, if they are temporarily
supplied with another entangled state $|\phi\rangle$, they can always
achieve the transformation from $|\psi_1\rangle$ to $|\psi_2\rangle$. The
state $|\phi\rangle$ is not consumed and can therefore be viewed as a {\it
catalyst} for this transformation}
\label{catal}
\end{center}
\end{figure}

The main result of this letter is the proof that entanglement-assisted
local transformations are indeed more powerful than ordinary local
transformations.
This result is significant in a number of ways. First of all, 
it provides a concrete mechanism by which
Alice and Bob can enhance their entanglement-manipulation ability. For
example, we will demonstrate that entanglement concentration is more
efficient with ELQCC than with only LQCC. Moreover, the definition of a
meaningful new class of entanglement transformations demonstrates that the
structure of entanglement, even for pure, bipartite states, is still not 
completely understood.

Let us begin then with an explicit example of the power of
entanglement-assisted transformations. The central tool we will require for
this is Nielsen's theorem \cite{Nielsen,Jonathan1}.

{\bf Theorem (Nielsen): }Let $\left| \psi_1 \right\rangle =\sum^{n}_{i=1}
\sqrt{\alpha_i}\left|i_A \right\rangle \left|i_B \right\rangle$ and $\left|
\psi_2 \right\rangle =\sum^{m}_{i=1}\sqrt{\alpha'_i}\left|i_A \right\rangle
 \left|i_B \right\rangle $ be pure bipartite states, with Schmidt
coefficients  \cite{Schmidt1} respectively $\alpha_1\geq...\geq\alpha _n >0$ and
$\alpha'_1\geq...\geq\alpha'_m >0$ (we can refer to such distributions as
``ordered Schmidt coefficients'', or OSCs). Then a transformation $T$ that
converts $\left|\psi_1\right\rangle $ to $\left|
\psi_2 \right\rangle $ with $100\%$ probability can be realised using LQCC
iff the OSCs $\left\{\alpha_i \right\}$ are {\em majorized} \cite{Olkin1} by
$\left\{\alpha'_i\right\}$, that is, iff for $1\leq l\leq n$

\begin{equation}
\sum_{i=1}^{l}\alpha _{i}\leq \sum_{i=1}^{l}\alpha _{i}^{\prime }.
\end{equation}

One consequence of Nielsen's theorem is that there exist pairs $\left|
\psi_1 \right\rangle $ and $\left| \psi_2 \right\rangle $ where neither state
is convertible into the other with certainty under LQCC. Such pairs are
called {\em incomparable} \cite{Nielsen}, and can be indicated by
$\left|\psi_1 \right\rangle \nleftrightarrow \left|
\psi_2 \right\rangle$. Examples are the following two states:
\begin{eqnarray}
\left| \psi _{1}\right\rangle &=& \sqrt{0.4} \left| 00\right\rangle +%
\sqrt{0.4}\left| 11\right\rangle +\sqrt{0.1}\left|22\right\rangle
+\sqrt{0.1}\left| 33\right\rangle  \nonumber \\
\left| \psi _{2}\right\rangle &=&\sqrt{0.5}\left| 00\right\rangle +
\sqrt{0.25} \left| 11\right\rangle +\sqrt{0.25} \left| 22\right\rangle.
 \label{example}
\end{eqnarray}
It can easily be checked that $\alpha_1<\alpha'_1$ but $\alpha_1 + \alpha_2
> \alpha'_1 + \alpha'_2$, so indeed $\left|\psi_1 \right\rangle \nleftrightarrow
\left| \psi_2 \right\rangle$. If Alice and Bob share one of these states and
wish to convert it to the other using LQCC, they must therefore run the
risk of failure. Their greatest probability of success is given by
\cite{Vidal1}
\begin{equation}
p_{\max}\left(\left|\psi_1\right\rangle \rightarrow
\left|\psi_2\right\rangle \right)= \left. \min \right|_{1\leq l\leq n}
\frac{E_l\left(\left|\psi_1\right\rangle \right)} {E_l\left(\left|\psi_2\right
\rangle \right)}
\label{prob}
\end{equation}
where
$E_l\left(\left|\psi_1\right\rangle\right)=1-\sum_{i=1}^{l-1}\alpha_i$. For
instance, in the case of the pair in Eq. (\ref{example}), $p_{\max}$ is
only 80\%.

Suppose now that Scrooge lends them the 2-qubit state
\begin{equation}
\left| \phi \right\rangle =\sqrt{0.6}\left| 44\right\rangle
+\sqrt{0.4}\left| 55\right\rangle.
\label{catalyst}
\end{equation}
The Schmidt coefficients $\gamma _{k}$, $\gamma'_{k}$ of the product states
$\left| \psi
_{1}\right\rangle   \left| \phi \right\rangle ,$ $\left| \psi
_{2}\right\rangle   \left| \phi \right\rangle $, given in decreasing
order, are
\begin{eqnarray}
\left| \psi _{1}\right\rangle   \left| \phi \right\rangle
&:&0.24,0.24,0.16,0.16,0.06,0.06,0.04,0.04, \\
\left| \psi _{2}\right\rangle   \left| \phi \right\rangle
&:&0.30,0.20,0.15,0.15,0.10,0.10,0.00,0.00.  \nonumber
\end{eqnarray}
so that $\sum_{i=1}^{k}\gamma _{k} \leq \sum_{i=1}^{k}\gamma' _{k}
,1\leq k\leq 8$. Nielsen's theorem then implies that
the transformation $\left| \psi_{1}\right\rangle   \left| \phi
\right\rangle \rightarrow \left| \psi_{2}\right\rangle
\left| \phi \right\rangle$ can, in fact, be realised with 100\%
certainty using LQCC. Alice and Bob can complete their task and still
return the borrowed state $\left|\phi \right\rangle $ to Scrooge. This
state acts therefore much like a {\em catalyst} in a chemical reaction: its
presence allows a previously forbidden transformation to be realised, and
since it is not consumed it can be reused as many times as desired. This
represents a fundamental distinction between the present effect and
previous proposals for using entanglement as an enhancing factor, such as
entanglement pumping \cite{Cirac} or the activation of bound entanglement
\cite{Horodeckis1}, where the extra entanglement was either used up or
transformed. We shall thus adopt the ``catalysis'' metaphor as a convenient
way of referring to our novel effect.

Nielsen's theorem, along with its generalisation given in
\cite{Jonathan1}, provides a complete answer to the question ``which
transformations on a pure bipartite state are possible under LQCC?'' It
would, of course, be desireable to find analogous conditions for ELQCC.
For instance, given $\left|\psi_{1}\right\rangle$,$
\left|\psi_{2}\right\rangle$, we would like to know when there exists
an appropriate catalyst state $\left|\phi \right\rangle$. Mathematically,
this means that given the OSCs $\left\{ \alpha_i \right\},\left\{ \alpha'_i
\right\}$, we have to determine when there exists other OSCs $\left\{
\beta_k \right\}$, such that $\left\{ \alpha_i \beta_k \right\}$ is majorized
by $\left\{ \alpha'_j \beta_k \right\}$. Unfortunately, this problem seems
in general to be hard to solve analytically \cite{Olkin2}. The difficulty
lies in the fact that, before Nielsen's theorem can be applied to the
tensor products, their Schmidt coefficients must be {\em sorted } into
descending order. No general analytic way for doing this is known, so we
are at present confined to searching numerically for appropriate catalysts.
Nevertheless, it is possible to present a few interesting partial results.

{\bf Lemma 1:} No transformation can be catalysed by a maximally entangled
state $\left|\varphi_p\right\rangle = (1/\sqrt{p})\sum_{i=1}^{p}
\left|i_A\right>\left|i_B\right\rangle$.

{\em Proof:} The Schmidt coefficients $\gamma_j$ of
$\left|\psi_1\right\rangle\left|\varphi_p\right\rangle$ are just
$\frac{\alpha_i}{p}$, each one being $p$-fold degenerate. In this case
sorting them is trivial, and we can write that, for any $l$, 
$\sum_{j=1}^{pl}\gamma_j =\sum_{i=1}^{l}\alpha_i$. Now, by Nielsen's 
theorem, if $\left|\psi_1\right\rangle \nrightarrow
\left|\psi_2\right\rangle$ under LQCC, then for some 
$l=l_0$ we have $\sum_{i=1}^{l_0}\alpha_i >
\sum_{i=1}^{l_0}\alpha'_i \Rightarrow \sum_{j=1}^{pl_0}\gamma_j >
\sum_{j=1}^{pl_0}\gamma'_j \Rightarrow \left|\psi_1\right\rangle
\left|\varphi_p\right\rangle\nrightarrow \left|\psi_2\right\rangle
\left|\varphi_p\right\rangle$ under LQCC $\square$

This result shows a surprising property of catalysts: they must be partially 
entangled. Roughly speaking, if the catalyst has ``not enough" entanglement,
Alice and Bob will not be able to transform $\left|\psi_{1}\right\rangle$ 
into $\left|\psi_{2}\right\rangle$ with certainty, but if it has 
``too much" then they will not be able to return it intact to Scrooge.

 {\bf Lemma 2: } Two states are {\em interconvertible} (i.e., both
$\left|
\psi_{1}\right\rangle \rightarrow\left| \psi _{2}\right\rangle $
and $\left| \psi _{2}\right\rangle \rightarrow \left| \psi
_{1}\right\rangle $) under ELQCC iff they are equivalent up to local unitary
transformations.

{\it Proof: }Suppose that $\left| \psi
_{1}\right\rangle  \leftrightarrow \left| \psi _{2}\right\rangle $ under ELQCC .
Then there exist $\left| \eta \right\rangle ,\left| \phi \right\rangle $
such that both $ \left| \psi _{1}\right\rangle \left| \phi \right\rangle
\rightarrow \left| \psi _{2}\right\rangle \left| \phi \right\rangle $ and $\left| \psi
_{2}\right\rangle \left| \eta \right\rangle \rightarrow \left| \psi
_{1}\right\rangle \left| \eta \right\rangle $ are possible under LQCC. This
means that $\left| \psi _{1}\right\rangle \left|\phi \right\rangle
\left|\eta \right\rangle$ and $ \left| \psi _{2}\right\rangle \left|
\phi \right\rangle \left| \eta \right\rangle $ are interconvertible
under LQCC, which happens iff their Schmidt coefficients are identical
\cite{Vidal2,Nielsen}. This in turn implies that the Schmidt coefficients
of $\left| \psi
_{1}\right\rangle $ and $\left| \psi _{2}\right\rangle $ are also identical,
and thus that they are equivalent under local unitary rotations $\square$

One consequence of Lemma 2 is that if a transition that is forbidden under
LQCC can be catalysed by extra entanglement (i.e.
$\left|\psi_{1}\right\rangle
\nrightarrow \left| \psi _{2}\right\rangle $ under LQCC but $\left| \psi_{1}\right\rangle
\rightarrow \left| \psi _{2}\right\rangle $ under ELQCC), then
the reverse transition (from $\left| \psi_{2}\right\rangle$ to $\left|\psi
_{1}\right\rangle $) must be impossible even under ELQCC. In particular,
only transitions between incomparable states may be catalysed. Therefore,
catalysis is impossible if $\left|\psi_{1}\right\rangle $ and $\left|
\psi_{2}\right\rangle $ are both $2\times 2$ states, for in this case it is
always true that either $\left|
\psi_{1}\right\rangle \rightarrow \left|\psi _{2}\right\rangle $ or
$\left| \psi _{2}\right\rangle \rightarrow \left|
\psi_{1}\right\rangle $ under LQCC.

A somewhat more surprising result is that catalysis is also impossible when
$\left| \psi _{1}\right\rangle $ and $\left| \psi _{2}\right\rangle $ are
both $3\times 3$ states. In this case incomparable states do exist \cite
{Nielsen}, so Lemma 2 does not immediately rule it out. To see that it
actually cannot occur, we must look more closely at the relevant Schmidt
coefficients.

{\bf Lemma 3: }Let $\left| \psi _{1}\right\rangle ,\left| \psi
_{2}\right\rangle \,$ be $n \times n$-level states, with OSCs $\{\alpha _{i}\},\{\alpha
_{i}^{\prime }\},1\leq i\leq n$. Then $ \left| \psi _{1}\right\rangle
\rightarrow \left| \psi _{2}\right\rangle $ under ELQCC only if both
\begin{equation}
\alpha _{1} \leq \alpha _{1}^{\prime } \;,\;
\alpha _{n} \geq \alpha _{n}^{\prime }. \label{lem2a}
\end{equation}
{\it Proof: }Let $\left\{\beta _{j}\right\}_{j=1}^m$ be the OSCs of $\left|
\phi\right\rangle $. Then the largest and smallest
Schmidt coefficient of $\left| \psi_{1}\right\rangle \left| \phi
\right\rangle $ are, respectively, $\gamma _{1}=\alpha _{1}\beta _{1}$ and $
\gamma _{nm}= \alpha_{n}\beta _{m}$ 
(analogous expressions hold for $\left| \psi_{2}\right\rangle
\left|\phi \right\rangle $ ). Nielsen's theorem now tells us that, if $\left| \psi
_{1}\right\rangle \left| \phi \right\rangle \rightarrow \left| \psi
_{2}\right\rangle \left| \phi \right\rangle $ under LQCC, then $\gamma
_{1}\leq \gamma _{1}^{\prime }$ and $\sum_{k=1}^{nm-1}\gamma _{k}=1-\gamma
_{nm}\leq \sum_{k=1}^{nm-1}\gamma _{k}^{\prime }=1-\gamma _{nm}^{\prime }$,
from which Eq. (\ref{lem2a}) follows $\square $

Suppose now that $\left| \psi _{1}\right\rangle $ and $\left| \psi
_{2}\right\rangle $ are incomparable $3\times 3$ states. Then Nielsen's
theorem implies that one of two possibilities must hold: either
\begin{equation}
\left\{
\begin{array}{c}
\alpha _{1}>\alpha _{1}^{\prime } \\
\alpha _{1}+\alpha _{2}<\alpha _{1}^{\prime }+\alpha _{2}^{\prime }
\end{array}
\right. \text{or }\left\{
\begin{array}{c}
\alpha _{1} < \alpha _{1}^{\prime } \\
\alpha _{1}+\alpha _{2}>\alpha _{1}^{\prime }+\alpha _{2}^{\prime }
\end{array}
\right. .
\end{equation}
In either case, Eq. (\ref{lem2a}) is violated, so $\left|
\psi _{1}\right\rangle {\nleftrightarrow}\left| \psi _{2}\right\rangle $
under ELQCC. In other words, there are pairs of states which are
incomparable even in presence of extra entanglement.

In the $4\times 4$ case, we have seen by example (Eq. (\ref{example}))\
that catalysis is indeed possible. Lemma 3 shows that the {\it only }case
where it can happen is when the following conditions are all satisfied
\begin{equation}
\alpha _{1} \leq \alpha _{1}^{\prime }  \label{4x4a}, \;\;
\alpha _{1}+\alpha _{2} >\alpha _{1}^{\prime }+\alpha _{2}^{\prime },\;\;
\alpha _{4} \geq \alpha _{4}^{\prime }.
\end{equation}
where the second condition ensures that the transformation is not possible 
under LQCC alone. Indeed, the states $|\psi_{1}\rangle,|\psi_{2}\rangle$
in Eq. (\ref{example}) are of this type.

The concept of entanglement-assisted transformations may be extended in a
number of ways. An example is when the presence of a catalyst state does
not allow Alice and Bob to transform $\left| \psi
_{1}\right\rangle$ into $\left| \psi _{2}\right\rangle$ with certainty, but
still increases the {\em optimal probability} with which this can be done.
For instance, consider the incomparable $3\times3$ states $\left|
\psi_{1}\right\rangle=\sqrt{0.5}\left|00\right\rangle+\sqrt{0.4}\left|11\right\rangle+
\sqrt{0.1}\left|22\right\rangle$ and $\left| \psi _{2}\right\rangle=\sqrt{0.6}
\left|00\right\rangle+\sqrt{0.2}\left|11\right\rangle+
\sqrt{0.2}\left|22\right\rangle$. From eq. (\ref{prob}) the
optimal probability of converting $\left| \psi
_{1}\right\rangle$ into $\left| \psi _{2}\right\rangle$ under LQCC is $80 \%$, and
Lemma 3 tells us that this cannot be increased to 100\% by the use of any
catalyst. Nevertheless, Eq. (\ref{prob}) also shows that $p_{\max}\left(
\left| \psi_{1}\right\rangle \left|\phi \right\rangle \rightarrow \left|
\psi _{2}\right\rangle \left| \phi \right\rangle \right)$ can be as large as $90.04\%$
when $\left|\phi\right\rangle =
\sqrt{0.65}\left|33\right\rangle+\sqrt{0.35}\left|44\right\rangle$.

Even this limited enhancement is not always possible, as shown by the
following result:

{\bf Lemma 4:} Let $\left| \psi _{1}\right\rangle, \left| \psi
_{2}\right\rangle$ be $n\times n$ bipartite states with OSCs $\{\alpha _{i}\},\{\alpha
_{i}^{\prime }\}$, and such that $p_{\max}\left(
\left| \psi _{1}\right\rangle \rightarrow \left| \psi_{2}\right\rangle \right)$
under LQCC is $\frac{\alpha_n}{\alpha'_n}$. Then this probability cannot be
increased  by the presence of {\em any} catalyst state.

{\em Proof:} Let $\left| \phi\right\rangle$ be a bipartite state with
OSCs $\left\{\beta\right\}_{i=1}^m. $
From Eq. (\ref{prob}), the optimal probability of converting
$\left|\psi_{1}\right\rangle \left| \phi \right\rangle $ into $\left| 
\psi_{2}\right\rangle  \left| \phi \right\rangle $ under LQCC is given by
\begin{equation}
p_{max}= \min_{l\leq nm} \frac{E_l \left( \left|
\psi_1\right\rangle   \left| \phi \right\rangle \right) } {E_l \left(\left|
\psi_2 \right\rangle    \left| \phi \right\rangle \right)}
\leq \frac{E_{nm} \left( \left| \psi_1 \right\rangle   \left|
 \phi \right\rangle \right) }{E_{nm} \left(\left|  \psi_2 \right\rangle
  \left| \phi\right\rangle \right)}
= \frac{\alpha_n}{\alpha'_n},
\end{equation}
where we have used that $E_{nm}(\left|\psi_1\right\rangle\left|\phi\right\rangle)
=\alpha_n\beta_m$ $\square$

In particular, Lemma 4 applies when $\left| \psi _{1}\right\rangle$ 
has $n$ Schmidt coefficients and $\left| \psi _{2}\right\rangle$ is the maximally
entangled state  $\left|\varphi_n\right\rangle$, for in this case $p_{\max} \left(
\left| \psi _{1}\right\rangle \rightarrow \left| \varphi_{n}\right\rangle \right) =$$
n\alpha_n$ \cite{Lo1,Vidal1}. 
At first sight, this may seem to indicate 
that catalytic effects cannot increase the efficiency with which entanglement 
can be concentrated into maximally-entangled form. It turns
out, however, that the opposite is actually the case. To see this, recall
first that an {\em entanglement concentration protocol} (ECP) can be
defined \cite{Bennett96a,Jonathan1} as any sequence of LQCC's that
transform a given partially entangled state $\left|\psi_1\right\rangle$
into a maximally entangled state $\left|\varphi_m \right\rangle$ of $m$
levels, with probability $p_m$ (note that $\left|\varphi_1 \right\rangle$
is a disentangled state). Among all these protocols, the {\em optimal} is
the one that yields on average the greatest amount of concentrated
entanglement i.e., that maximises $\left<E\right>=\sum_{m=1}^n p_m\ln m$
over all possible distributions $\{p_m\}$ compatible with LQCC. The maximum
value $\left<E\right>_{max}$ is the (finite) distillable entanglement of
$\left|\psi_1\right\rangle$ \cite{note}. In \cite{Jonathan1,Hardy1} it was
shown that
$
\left<E\right>_{max}= \sum_{m=0}^n m\ln m(\alpha_m-\alpha_{m+1}),
\label{purif}
$
corresponding to a probability distribution $p_m^{opt} = m(\alpha_m
-\alpha_{m+1})$, where $\left\{\alpha_i\right\}$ are the OSCs of
$\left|\psi_1\right\rangle$ and $a_{n+1}=0$.
\begin{figure}[hbt]
\begin{center}
\leavevmode
\epsfxsize=6cm
\epsfbox{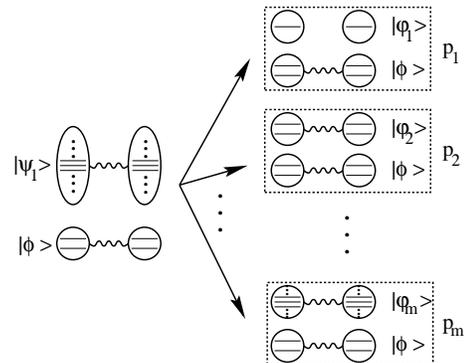}
\caption{\narrowtext A catalysed entanglement concentration protocol. Alice and
Bob share a state  $\left|\psi_1\right\rangle   \left|\phi\right\rangle$
and by LQCC convert it with probability $p_m$ into a product
$\left|\varphi_m\right\rangle
\left|\phi\right\rangle $ between a maximally entangled state of $m$ levels
and the catalyst state $\left|\phi\right\rangle$. The entanglement in
$\left|\psi_1\right\rangle$ may be concentrated with greater efficiency in
this way than in the absence of $\left|\phi\right\rangle$ }
\label{catECP}
\end{center}
\end{figure}
A {\em catalysed} ECP (Fig. 2) is then any sequence of LQCCs that transform
the product $\left|\psi_1\right\rangle \left|\phi\right\rangle$ (for some
catalyst state $\left|\phi\right\rangle$) into one of the states
$\left|\varphi_m\right\rangle\left|\phi\right\rangle $, with probability
$p_m$. It turns out that in this case the distillable entanglement
$\left<E\right>_{\max} (\left|\phi\right\rangle)$ can be {\em larger} than the
value given above. To show this, we use a general technique 
for optimising entanglement transformations, presented in \cite{Jonathan1}.
From the generalised Nielsen's theorem \cite{Jonathan1}, a catalysed ECP
with probability distribution $p_m$ can be realised using LQCCs iff the
following constraints are satisfied for $1\leq l \leq n$
\begin{equation}
\sum_{m=1}^{n}p_m E_l\left(\left|\varphi_m\right\rangle   \left|\phi\right\rangle
\right) \leq E_l \left( \left|\psi_1\right\rangle   \left|\phi\right\rangle \right),
\end{equation}
where $E_l$ is the same as in Eq. (\ref{prob}). The optimal protocol can
then be found by maximizing $\left<E\right>(\left|\phi\right\rangle )$ with
respect to $p_m$, given these constraints. This is a standard {\em linear
programming} problem \cite{Walsh}, for which an exact solution can always
be found in any particular case. In Fig. 3 we plot $\left<E\right>_{\max}
(\left|\phi\right\rangle )$ for the case where $\left|\psi_1\right\rangle$
is a $3\times3$ state with Schmidt coefficients $\alpha_1=0.5$,
$\alpha_2=0.3$, $\alpha_3=0.2$, and where $\left|\phi\right\rangle$ is a
a $2\times2$ state. We can see that some catalysts allow a substantial
increase in the entanglement yield relative to the one achievable using
only LQCC.

How does this happen, even under the constraints implied by Lemma 4? It
turns out that, although Lemma 4 forbids $p_n$ from increasing in the
presence of a catalyst, the same is not true for $p_{n-1}$. For instance,
in the example above the optimal probability distribution without a
catalyst (i.e., with a disentangled catalyst) is \cite{Jonathan1}
$p_3^{opt}=0.6, p_2^{opt}=p_1^{opt}=0.2$. On the other hand, in the
presence of a catalyst with Schmidt components
$\beta_1=0.5825,\beta_2=.4175$, it becomes $p_3^{opt}=0.6,
p_2^{opt}=0.3178, p_1^{opt}=0.0822$. Effectively, the presence of a catalyst
allows us to syphon probability away from the unwanted outcome where all
the entanglement is lost and into one where a maximally entangled state is
obtained.
\begin{figure}[hbt]
\begin{center}
\leavevmode
\epsfxsize=7cm
\epsfbox{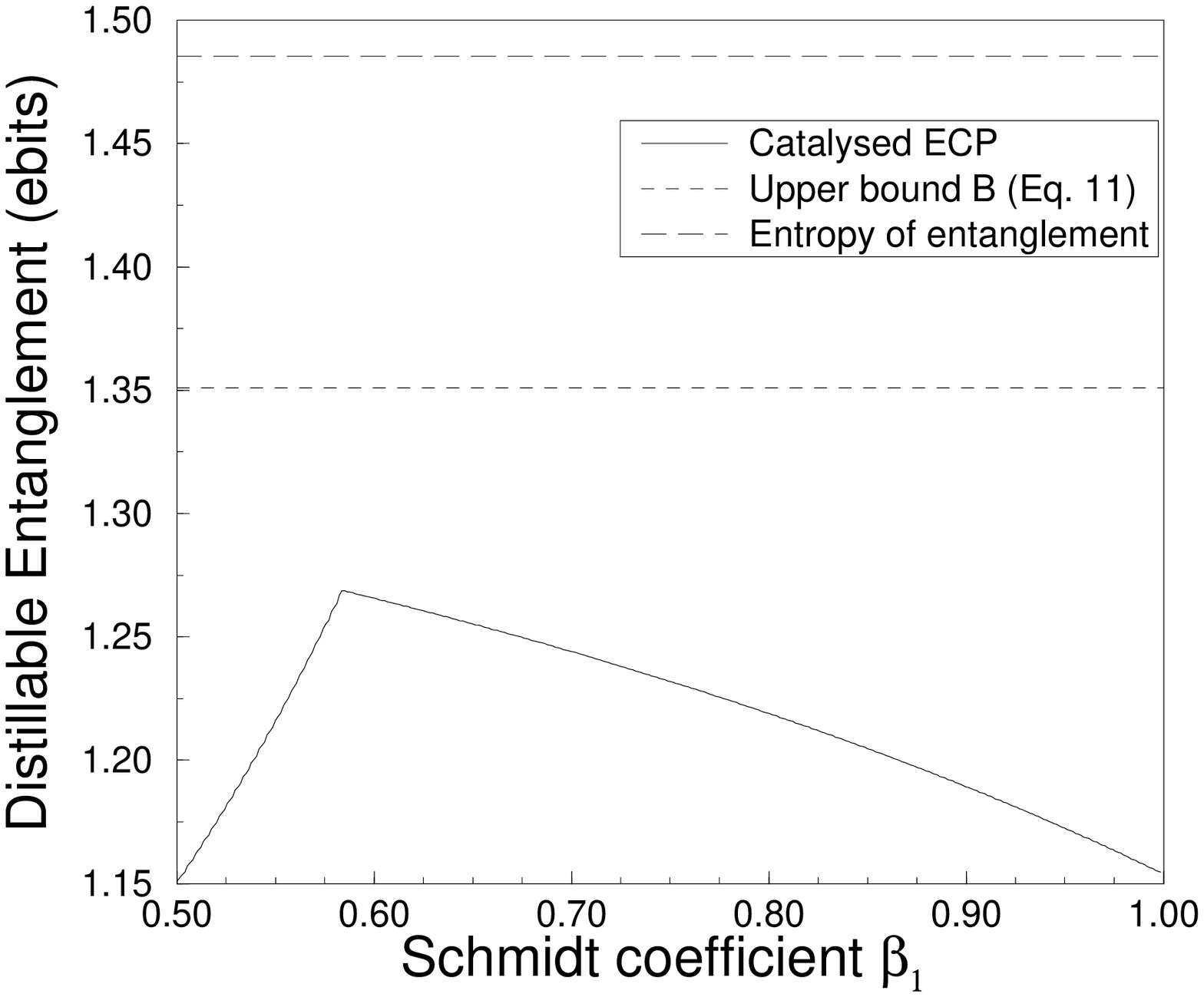}
\caption{\narrowtext Finite distillable entanglement $\left<E\right>_{\max} 
(\left|\phi\right\rangle)$ (in ebits) of the $3\times3$ state $\left|\psi_1\right\rangle=
\sqrt{0.5}\left| 00\right\rangle + \sqrt{0.3} \left| 11\right\rangle 
+\sqrt{0.2} \left| 22\right\rangle$, in the presence of a catalyst state 
$\left|\phi\right\rangle$ with Schmidt coefficients $\beta_1 \geq \beta_2$. 
The horizontal lines are upper bounds to $\left<E\right>_{\max} 
(\left|\phi\right\rangle )$ given by Eq. (12) and by the entropy of 
entanglement of $|\psi_1\rangle$. Note that neither product states nor
maximally entangled states are useful as catalysts.}
\label{distill}
\end{center}
\end{figure}
How far can this enhancing effect be used to increase the distillable
entanglement $\left<E\right>_{max}(|\phi\rangle)$ ? Lemma 4 gives us 
immediately the following upper bound
\begin{equation}
 B \equiv  n\alpha_n \ln n + (1-n\alpha_n)\ln (n-1) \ge \left<E\right>_{max}(|\phi\rangle)
\end{equation}

This simply corresponds to a case where $p_n$ is maximum, and all the
remaining probability is assigned to obtaining $\left|\varphi_{n-1}\right\rangle$.
 Another upper bound is the {\em asymptotic} distillable entanglement per copy 
\cite{note}. These bounds are unrelated: there are states like the one in Fig. 
\ref{distill}, for which $B<S$, and others for which $S<B$. It is an open question 
whether any of these bounds can in general be reached by the use of an appropriate 
catalyst.

To summarise: we have presented a counter-intuitive effect by means of which
local entanglement transformations of finite states may be catalysed in
the presence of `borrowed' entanglement. Our results raise many interesting questions. For
instance, what are {\em sufficient} conditions for the existence of catalysts? Are catalysts
always more efficient as their dimension increases? We hope that the intricacy of 
this effect may convince readers that there is more to pure-state entanglement than just asymptotic
properties, and that no single ``measure'' of entanglement can fully capture it all.

Acknowledgements: We would like to thank I. Olkin, A. Uhlmann and
O. Pretzel for helpful comments on majorization and tensor products
and especially S. Popescu for inspiring discussions. We
acknowledge the support of the Brazilian agency Conselho Nacional de
Desenvolvimento Cient\'{\i}fico e Tecnol\'{o}gico (CNPQ), the ORS Award Scheme,
the United Kingdom Engineering and Physical Sciences Research Council,
the Leverhulme Trust, the European Science Foundation, and the European Union.

\end{multicols}

\begin{references}
\bibitem{Schmidt1} Note that the `Schmidt coefficients' defined
here are the squares of the usually defined Schmidt coefficients.
\bibitem{Olkin2} I. Olkin and A. Uhlmann, private communications.
Indeed, there does not seem to have been any extensive work by
mathematicians on the subject of majorization and tensor products.
\end{references}
\end{document}